\documentclass[12pt]{iopart}

\usepackage{graphicx}
\usepackage{color}
\usepackage{array}
\usepackage{cite}

\begin{document}

\paper[Phenomenon-based prediction of RBE of ion beams by means of the MSA]
{Phenomenon-based prediction of relative biological effectiveness of ion beams by means of the MultiScale Approach}

\author{A. Verkhovtsev$^{1,2,3}$, E. Surdutovich$^4$, A.V. Solov'yov$^{2,3}$}

\address{$^1$ Department of Medical Physics in Radiation Oncology, German Cancer Research Center (DKFZ),
Im Neuenheimer Feld 280, 69120 Heidelberg, Germany}
\address{$^2$ MBN Research Center, Altenh\"oferallee 3, 60438 Frankfurt am Main, Germany}
\address{$^3$ Ioffe Institute, Politekhnicheskaya 26, 194021 St. Petersburg, Russia}
\address{$^4$ Department of Physics, Oakland University, Rochester, Michigan 48309, USA}

\ead{a.verkhovtsev@dkfz-heidelberg.de}

\date{\today}

\begin{abstract}
Relative biological effectiveness (RBE) is a key quantity for the description of radiobiological effects
induced by charged-particle irradiation in the context of ion-beam cancer therapy.
Since RBE is a complex function that depends on different physical, chemical and biological parameters,
a fundamental understanding of these effects becomes increasingly important for clinical applications.
The phenomenon-based MultiScale Approach to the physics of radiation damage with ions (MSA)
provides a tool for a molecular-level understanding of physical and chemical mechanisms of
radiation biodamage and allows to quantify macroscopic biological effects caused by ion irradiation.
This study reports the first application of the MSA for the analysis of RBE of ion beams.
As a case study, we quantify the response of human normal tissue cell lines to carbon-ion
irradiation at different values of linear energy transfer.
RBE corresponding to different biological endpoints as well as other radiobiological parameters,
such as clonogenic cell survival as a function of dose and inactivation cross section,
are analyzed and compared with experimental data.
Good agreement with an extensive data set emphasizes predictive power of the MSA.
The method is also used to make predictions of RBE at high values of LET where RBE decreases
due to the ``overkill'' effect.
In this regime, the dose needed to achieve a given biological effect is deposited by a few ions.
The fact that a given number of ions may produce more damage than needed for a given effect
leads to a significant (up to 20\%) variation in RBE.
This effect should be taken into consideration in the analysis of experimental data on irradiation
with high-LET ions.
\end{abstract}


\maketitle

%%%%%%%%%%%%%%%%%%%%%%%%%%%%%%%%%%%%%%%%%%%%%%%%%%%%%%%%%%%%%%%%%%%%%
%%%%%%%%%%%%%%%%%%%%%%%%%%%%%%%%%%%%%%%%%%%%%%%%%%%%%%%%%%%%%%%%%%%%%

\section{Introduction}

%%%%% 1) IBCT, LET

Ion-beam cancer therapy (IBCT), also known as hadron therapy,
is an emerging, rapidly developing treatment technique \cite{Schardt_2010_RevModPhys.82.383,
Jakel2008_MedPhys.35.5653, Loeffler_2013_NatRevClinOncol.10.411}.
IBCT provides advances in cancer treatment due to the possibility of high dose localization in the tumor region.
This allows to maximize cell killing within the tumor whilst simultaneously minimizing the radiation damage
to surrounding healthy tissue.
The advantages of IBCT over conventional radiotherapy with photons stem from the fundamental
difference between the energy deposition profiles for heavy charged projectiles and photons
\cite{Schardt_2010_RevModPhys.82.383, Surdutovich_2014_EPJD.68.353}.
The energy deposited by photons rises at shallow penetration depths and then decreases exponentially
as photons are absorbed \cite{Mohan_2017_AdvDrugDelRev.109.26}.
The profile for ions is characterized by a plateau region and the Bragg peak, that is a sharp maximum
in the depth-dose curve close to the end of ions' trajectories \cite{Hall_Giaccia_2018}.

%%%%% 2) RBE

There is an abundance of experimental evidence that irradiation with energetic ion beams
results in enhanced cell killing as compared to photon irradiation at the same dose.
In order to account for this effect the concept of relative biological effectiveness (RBE)
has been introduced \cite{IAEA_TRS461_RBE}.
RBE is defined as the ratio of a dose of photons to a dose of ions
(or, in general, of any other radiation modality) leading to the same biological effect,
\begin{equation}
{\rm RBE} = \frac{d_{\rm ph}}{d_{\rm ion}} \ .
\end{equation}
This expression allows one to calculate, for a given ion dose absorbed, the isoeffective photon dose
and thus to estimate the biological effect of ion irradiation on the basis of the
well-known response to a reference photon beam \cite{Karger_2018_PMB.63.01TR02}.
Despite being a simple concept, RBE depends on many physical
(radiation type, energy, linear energy transfer (LET), radiation dose, dose rate, fractionation scheme, etc.),
chemical (e.g., oxygen concentration in the target) and biological (biological endpoint,
intrinsic radiosensitivity of a given cell line, cell cycle phase, proliferation rate, etc.) parameters
\cite{Surdutovich_2014_EPJD.68.353, Karger_2018_PMB.63.01TR02, Paganetti_2014_PMB.59.R419}.

In principle, any biological endpoint may be used to determine RBE.
The effects of ion beams have been studied mostly in biological systems \textit{in vitro} with
clonogenic cell survival being a commonly used endpoint.
In this case, irradiations with photons and ions are considered to be isoeffective if the
dose-dependent survival fractions measured in the clonogenic assay are the same.

%%%%% 3) calculation of RBE, RB models

Different radiobiological models have been developed to describe experimental outcomes and understand
how physical parameters of irradiation impact the biological response of cells and tissues.
The most widely known approaches are the Local Effect Model (LEM)
\cite{Schardt_2010_RevModPhys.82.383, Scholz_1997_REB.36.59, Elsasser_2008_IJROBP.71.866, Friedrich_2012_IJRB.88.103}
that is used for treatment planning in ion-beam centers in Europe,
Microdosimetric Kinetic Model (MKM) \cite{Hawkins_1996_IJRB.69.739, Hawkins_2003_RadiatRes.160.61}
as well as the modified MKM (MMKM) \cite{Inaniwa_2010_PMB.55.6721, Kase_2011_JRadiatRes.52.59}
which is used clinically in Japan.
The LEM describes biological effects of ion beams on the basis of amorphous track structure in
combination with the known dose response curves for photon radiation.
The MKM and MMKM rely on microdosimetric concepts and on the estimation of the stochastic energy
deposition into volumes of micrometer dimensions \cite{Kelleler_1985_chapter}.
In the above-mentioned approaches the radiobiological effect of ions is quantified
by means of an empirical linear-quadratic (LQ) model,
\begin{equation}
-\ln{\Pi} = \alpha d + \beta d^2 \ ,
\label{LQ_model}
\end{equation}
where $\Pi$ is a surviving fraction of cells exposed to a given dose of radiation $d$.
The coefficients $\alpha$ and $\beta$, which characterize the response of biological systems to radiation,
are usually determined by fitting the experimental data on clonogenic cell survival.

The above-mentioned models are currently used in clinical practice for the dose optimization
and treatment planning.
However, being based on an empirical equation~(\ref{LQ_model}), these models cannot answer
many questions concerning the molecular-level mechanisms of radiation damage with ions.
As a result, the fundamental scientific knowledge of the involved physical, chemical and biological
effects is, to a significant extent, missing \cite{Nano-IBCT_book, Bacarelli_2010_EPJD.60.1}.
The understanding of radiation biodamage on a fundamental quantitative level is necessary to bring IBCT
optimization and planning to a higher scientific level in order to improve the currently existing
protocols \cite{Nano-IBCT_book}.

%%%%% 4) MSA, main idea in brief

The empirical level of the earlier concepts has triggered formulation of the MultiScale Approach to the physics
of radiation damage with ions (MSA) \cite{Solov'yov_2009_PRE.79.011909, Surdutovich_2014_EPJD.68.353, Nano-IBCT_book}.
The MSA has been developed to construct an inclusive scenario of processes leading to radiation damage
with the ultimate goal of its quantitative assessment.
This approach joins the knowledge about production of secondary electrons and other
reactive species in the vicinity of ion's path, the transport of these species, and
cross section of interaction with DNA molecules to calculate the probability of
important lesions, such as double and single strand breaks (DSBs and SSBs) per unit
length of ion's path \cite{Surdutovich_2014_EPJD.68.353}.
Then, a criterion for lethality of damage is established.
It states that if two SSBs occur in the vicinity of a DSB the damage is defined as lethal.
Finally, the probability of production of a lethal lesion is obtained as a function
of ion fluence, number density of chromatin in a target cell nucleus, and the LET-dependent
cross section of lethal damage \cite{Surdutovich_2014_EPJD.68.353, Verkhovtsev_2016_SciRep.6.27654}.

A series of works reviewed in \cite{Surdutovich_2014_EPJD.68.353}
was devoted to different processes and aspects of radiation damage with ions.
In that work, a recipe for the assessment of biodamage was suggested
and it was the first phenomenon-based calculation of a cell survival curve for ions.
The predictability of cell survival by the MSA was successfully tested on
a variety of cell lines with different values of LET and oxygenation conditions \cite{Verkhovtsev_2016_SciRep.6.27654}.
Another recent achievement of the MSA is formulation of a recipe for solving a
problem of variable cell survival probability along the spread-out Bragg peak
\cite{Surdutovich_2017_EPJD.71.210}.

In this paper the MSA methodology is applied to evaluate RBE of ions beams.
As a case study, we analyze the response of human normal tissue cell lines to
carbon-ion irradiation at different values of LET.
RBE corresponding to different biological endpoints as well as other radiobiological parameters,
such as clonogenic cell survival as a function of dose and inactivation cross section,
are analyzed and compared with experimental data compiled in
the Particle Irradiation Data Ensemble (PIDE) database \cite{Friedrich_2013_PIDE}.
A good agreement with experimental results clearly illustrates the capability of the MSA to
quantitatively describe RBE for different biological endpoints as well as other radiobiological parameters.
Finally, the MSA is used to make predictions of RBE at high values of LET
(above 100~keV/$\mu$m) where RBE for carbon ions is known to decrease due to ``overkill'' effect
\cite{IBT_book_Linz}.
The fact that a given number of high-LET ions may produce more damage than needed
for a given biological effect leads to significant variation of RBE.
Normal cell lines are chosen as an illustrative case study because their proliferation is highly
organized as compared to tumor cells.
This allows us to test robustness of the MSA-based methodology and justify the choice of its key parameters,
e.g., the genome size which remains almost constant in normal cells but may
vary greatly in different tumor cells \cite{Kops_2005_NatRevCancer.5.773}.

%%%%%%%%%%%%%%%%%%%%%%%%%%%%%%%%%%%%%%%%%%%%%%%%%%%%%%%%%%%%%%%%%%%%%
%%%%%%%%%%%%%%%%%%%%%%%%%%%%%%%%%%%%%%%%%%%%%%%%%%%%%%%%%%%%%%%%%%%%%
\section{The MSA methodology}
\label{Methods}

The MSA is a phenomenon-based approach that aims at predicting macroscopic biological effects caused
by ion radiation on the basis of physical and chemical effects related to the ion--medium interactions
on a nanometer scale \cite{Solov'yov_2009_PRE.79.011909, Surdutovich_2014_EPJD.68.353, Nano-IBCT_book}.
The key phenomena and processes addressed by the MSA are ion stopping in the medium,
production of secondary electrons and free radicals as a result of ionization and excitation of the medium,
transport of these species,
the interaction of secondary particles with biomolecules,
the analysis of induced damage,
and the evaluation of the probabilities of subsequent cell survival.
A comprehensive description of different aspects of the MSA was given in earlier publications
\cite{Surdutovich_2014_EPJD.68.353, Nano-IBCT_book}.
In this section, the formalism used for evaluation of radiobiological effects within this
approach is briefly outlined.

The assessment of RBE for ions starts from the calculation of survival curves for a given
type of cells irradiated with a given type of ions at given conditions.
This requires establishing of the relation between physical effects and radiation damage.
In regard to irradiation with ions, the key assumption adopted in the MSA is that the leading
cause of cell inactivation is the complexity of nuclear DNA damage
\cite{Ward_1995_RadiatRes.142.362, Amaldi_2005_RepProgPhys.68.1861,
Malyarchuk_2009_DNARepair.8.1343}.

The criterion for lethality of damage suggested in \cite{Surdutovich_2014_EPJD.68.353}
is based on the well-established hypothesis that among different DNA lesions caused
by interaction with secondary electrons and other reactive species
(e.g., free radicals and solvated electrons) the multiple damaged sites with
sufficient complexity may not be repaired~\cite{Ward_1995_RadiatRes.142.362,
Sage_2011_MutatRes.711.123, Malyarchuk_2009_DNARepair.8.1343}.
In the formulated recipe for the assessment of biodamage,
it was postulated that a complex lesion combined of a DSB and at least two other simple
lesions such as SSBs within two DNA twists is lethal for a cell \cite{Surdutovich_2014_EPJD.68.353}.
In our previous study \cite{Verkhovtsev_2016_SciRep.6.27654} this criterion was successfully applied
for a number of cell lines.

The multiple damage sites contain several lesions, each of which is caused by
independent agents, such as secondary electrons, free radicals or solvated electrons
\cite{Surdutovich_2011_PRE.84.051918}.
The MSA calculates the probability of such a site to be formed at a distance $r$
from a given ion's path; then the space averaging is applied.
The formation of the site on a DNA molecule segment requires a number of agents to reach it.
Secondary electrons produced following the ion's passage propagate in the medium on the femtosecond
time scale \cite{Surdutovich_2015_EPJD.69.193}.
They react with the DNA molecule producing lesions such as SSBs, DSBs, base damages, etc.
For most of them, the typical distance range of $r$ is within several nanometers from the ion's path and
the diffusion mechanism describes their transport adequately.
Less abundant high-energetic $\delta$-electrons (which are kinematically allowed to form on a plateau in front of the Bragg peak)
may induce damage sites several hundreds of nanometers away from the ion's path.

Mechanisms of transport of reactive species such as free radicals and solvated electrons depend on the ion's LET. If the LET is relatively small, the reactive species are formed in rather small numbers on a picosecond time scale and diffuse away from the ion's path reaching their targets on the way. Their lifetime is limited by their interactions with each other and with other components of the medium and,
provided the number densities are small enough, can be rather long, up to $10^{-4}$~s~\cite{Sonntag_ChemRadBiol, Alpen_RadiatBiophys}. Such long times may largely increase the ranges of distances exposed to reactive species. However, as the reactive species diffuse out, their number density decreases and may fall below the minimum density required for the formation of a lethal lesion.
Such a condition of the required minimum number density becomes the limiting factor for the effective range of
reactive species propagation.

The condition of the required minimum number density is introduced as a logical consequence of the introduction of the criterion for lesion lethality and the understanding that the formation of complex lesion requires a certain number of agents. The introduction of this condition is natural in the framework of the MSA. As the criterion itself is understood better the condition can be correspondingly modified. This is another reason why the MSA raised the interest to physical mechanisms of formation of lesions such as DSBs trying to understand how many secondary electrons or reactive species are required for their production.

At higher values of LET, the reactive species are produced in larger numbers.
The high reaction rates for interactions of reactive species may lead to their annihilation and not allow them to leave a few-nm ion track. A different physics, namely the predicted ion-induced shock waves \cite{Surdutovich_2010_PRE.82.051915}, steps in the scenario of radiation damage. The collective radial flow induced by these waves carries the reactive species reducing their number densities and saving them from annihilation \cite{Surdutovich_2015_EPJD.69.193}. This process happens on a picosecond time scale,
and the radial range to which the reactive species can propagate is determined by the strength of the shock wave.
This effect is complex and can be studied by means of advanced reactive molecular dynamics simulations
\cite{deVera_2018_EPJD_radiochem}.

Analytical considerations show that the effective range of propagation of reactive species by the shock wave-induced collective flow
is linear in the first order with respect to LET \cite{Surdutovich_2017_EPJD.71.285}.
Indeed, as it was shown \cite{Surdutovich_2010_PRE.82.051915, Surdutovich_2014_EPJD.68.353}
the pressure on the front of the shock wave is given by
\begin{equation}
P(r) = \frac{1}{\gamma + 1} \frac{\beta^4}{2} \frac{S_e}{r^2} \ ,
\end{equation}
where $\gamma = C_P / C_V \approx 1.2$ is the heat capacity ratio for water molecules,
$\beta = 0.86$ is a dimensionless constant, and $r(t) \propto \sqrt{t}$ is the radius of the wave front.
As the shock wave propagates in the radial direction away from the ion's path, it causes a rarefaction
in its wake and a cylindrical cavity of the radius $r_{\rm in} < r$ is formed.
The radius of the wave front increases as the pressure drops; this happens until the force inside the cavity
(due to surface tension pressure $\sigma/r_{\rm in}$) equilibrates the tearing force \cite{Surdutovich_2017_EPJD.71.285}.
The condition for saturation of the radial propagation of the shock wave-induced collective flow
can be estimated by equating the pressure force acting on a fragment of the wave front and the force due to
surface tension on the inner surface \cite{Surdutovich_2017_EPJD.71.285},
\begin{equation}
\frac{1}{\gamma + 1} \frac{\beta^4}{2} \frac{S_e}{r^2} \, 2\pi r l
= \frac{\sigma}{r} \, 2 \pi r l \ ,
\label{eq:SW_cond}
\end{equation}
where $r$ is considered to be the same on the left- and right-hand sides since the thickness of the wave front is much smaller than $r$.
The hydrodynamic phase, roughly described by this equation, ends when the pressure becomes uniform again. As a result of this phase, the reactive species are expected to be more or less uniformly distributed within the range $R$. After the hydrodynamic equilibrium is achieved the reactive species propagate further due to diffusion mechanism, but this stage is only of interest to us if the hydrodynamic range is smaller than that given by the required minimum number density. The linear dependence of $R$ on LET can be derived from Eq.~(\ref{eq:SW_cond}), however the numerical value of $R$ from that equation depends on the choice of $\sigma$, which is an uncertain quantity at the medium conditions arising in the shock wave.
Comparison of this analysis with the molecular dynamics simulations \cite{deVera_2016_EPJD.70.183, deVera_2018_EPJD_radiochem} show that shock waves decay on much shorter distances than it follows from Eq.~(\ref{eq:SW_cond}) evaluated at the normal conditions. The range of propagation of reactive species by the shock wave and its dependence on LET are currently under more thorough investigation using the molecular dynamics simulations including chemical reactions~\cite{Sushko_2016_EPJD.70.12} following Ref. \cite{deVera_2018_EPJD_radiochem}.

The secondary electron contribution to the scenario has been understood better than that of reactive species and a part of this understanding is that the damage is done not by their number density but rather by a collection of their incidences or hits of a particular molecular target.
Then, the total average fluence, $F_e(r)$, or the number of electrons incident on a typical target multiplied by an average probability of producing a simple lesion like SSB per hit, $\Gamma_e$, gives the total average number of simple lesions produced at a distance $r$ from the path, ${\cal N}_{\rm e}(r)$. The number of secondary electrons incident on such a target is calculated as an integral of the flux of secondary electrons through the target, $\Phi_e(r,t)$, over time, where the integral is taken from zero to the time $t_1$ on a femtosecond scale until when the electrons can be treated as ballistic particles. At the larger time scales, remaining electrons become solvated and are treated together with other reactive species created in the medium.
The flux $\Phi_e(r,t)$ is obtained by solving a three-dimensional diffusion equation
\cite{Surdutovich_2014_EPJD.68.353, Surdutovich_2015_EPJD.69.193}.
Explicit analytical expressions for $\Phi_{\rm e}(r,t)$ and $F_{\rm e}(r)$ can be found in Ref. \cite{Surdutovich_2014_EPJD.68.353}.

Even though the reactive species transport is much more subtle and less understood at the moment,
it is possible to cast it in the same form as that of secondary electrons. Then the complete picture looks as
\begin{equation}
\hspace{-2.5cm}
{\cal N}(r)
= {\cal N}_{\rm e}(r) + {\cal N}_{\rm r}(r)
= \Gamma_{\rm e}F_{\rm e}(r) + \Gamma_{\rm r}F_{\rm r}(r)
= \Gamma_{\rm e} \int_0^{t_1} \Phi_{\rm e}(r,t) \, {\rm d}t + \Gamma_{\rm r} \int_0^{t_2} \Phi_{\rm r}(r,t) \, {\rm d}t \ ,
\label{eq_01b}
\end{equation}
where quantities with index ${\rm r}$ represent similar quantities for reactive species.
The time limit $t_2$ depends on the physics involved in the transport of reactive species such as the shock wave-induced collective flow followed by hydrodynamic relaxation and diffusion; $t_2$ can be on the picosecond or even the nanosecond scale depending on the LET.
However, in our approach we choose $t_2$ on the picosecond scale according to the aforementioned criterion of the formation of lethal lesions in the vicinity of the track due to the creation of sufficiently high density of reactive species.

If the transport of reactive species were understood better, we would not have to discuss the detail of LET dependence of quantities in Eq.~(\ref{eq_01b}) as the corresponding integrands would naturally decrease with time and the distance. However, since there is no sufficient understanding of this transport, we will assume a linear dependence of the range of propagation of reactive species on LET following from Eq.~(\ref{eq:SW_cond}),
and take up a conservative estimate of $R\approx 10$~nm for carbon ions at their Bragg peak \cite{Surdutovich_2014_EPJD.68.353}. Within this range, the density of the reactive species should be high enough to ensure the production of DNA lethal lesions as discussed above. Furthermore, following Refs. \cite{Surdutovich_2014_EPJD.68.353, Verkhovtsev_2016_SciRep.6.27654},
the average number of lesions due to reactive species at a distance $r$ from the path is taken to be
\begin{equation}
{\cal N}_{\rm r}(r) = {\cal N}_{\rm r} \, \theta(R(S_e) - r) \ ,
\label{N_r_LET}
\end{equation}
where $\theta$ is the Heaviside function and $R=10~{\rm nm}\times S_e/S_{e,CBP}$ with $S_{e,CBP}$ being the LET of carbon ions at their Bragg peak.
The value of ${\cal N}_{\rm r}$ was estimated as 0.08 from the comparison of the experimental results \cite{Dang_2011_EPJD.63.359}
for plasmid DNA dissolved in pure water and in a scavenger-rich solution.
Again, more work is needed to obtain more detailed dependencies of $N_r$ and $R$ on LET.

After ${\cal N}(r)$ is obtained, the probability of production of a lethal lesion at a distance $r$ from the path,
${\cal P}_{l}(r)$, can be calculated according to the criterion of lethality determined in Refs. \cite{Surdutovich_2014_EPJD.68.353, Verkhovtsev_2016_SciRep.6.27654},
\begin{equation}
{\cal P}_{l}(r) = \lambda \sum_{\nu = 3}^\infty
{\frac{\left[ {\cal N}(r) \right]^{\nu} }{\nu !}\exp{\left[-{\cal N}(r)\right]}} \ ,
\label{eq_01c}
\end{equation}
where $\nu$ is the number of simple lesions per cluster and ${\cal N}(r)$ is defined in Eq.~(\ref{eq_01b}).
The sum starts with $\nu=3$, which makes the minimum order of lesion complexity at a given site
equal to three.
The factor $\lambda$ is the probability that one of the simple lesions is converted to a DSB.
This implies that in the current model the DSBs occur via SSB conversion, but in principle
other mechanisms can also be taken into account \cite{Surdutovich_2012_EPJD.66.206}.
The introduction of this factor relies on experimental findings~\cite{Huels_2003_JACS.125.4467, Sanche_2005_EPJD.35.367}
that the DSBs caused by electrons with energies higher than about 5~eV happen in one hit.
In this case, the subsequent break in the second strand of the DNA is due to the action of debris
generated by the first SSB.
In the cited works it was shown that if a single electron causes an SSB, the same electron causes
a DSB with a probability of about $0.1 - 0.2$ of that to create an SSB.
The value $\lambda = 0.15$ was suggested earlier~\cite{Surdutovich_2014_EPJD.68.353}
and has been utilized in the analysis presented below.
As demonstrated in our earlier work \cite{Verkhovtsev_2016_SciRep.6.27654} and as shown below,
this criterion is valid for the description of a large set of normal and tumor cell lines.

Equation~(\ref{eq_01c}) represents the radial distribution of lethal lesions.
Integration of ${\cal P}_{l}(r)$ over the area perpendicular to the ion's path
gives the number of lethal lesions per unit length of the ion's trajectory,
\begin{equation}
\frac{{\rm d}N_{l}}{{\rm d}x} =
%n_{\rm s} \, \int\limits_0^{\infty} {\cal P}_{l}(r) \, 2\pi r \, {\rm d}r  =
n_{\rm s} \, \int\limits_0^{R} {\cal P}_{l}(r) \, 2\pi r \, {\rm d}r  =
n_{\rm s} \, \sigma(S_e) \ .
\label{eq_02}
\end{equation}
Here, $n_{\rm s}$ is the number density of chromatin which is
proportional to the ratio of DNA base pairs accommodated in the cell nucleus to the
nuclear volume, $n_{\rm s} \sim N_{\rm bp}/V_{\rm n}$ \cite{Verkhovtsev_2016_SciRep.6.27654}.
If the correct dependence of ${\cal N}_r$ on $r$ were known, the upper limit in Eq.~(\ref{eq_02}) would not be necessary since the integrand would have plunged to zero at a certain distance.
In this work the upper limit corresponds to the range $R$ in the Heaviside function.
This is a combination of ranges of transport by the collective flow, hydrodynamic relaxation, and the diffusion of radicals.
The limiting factor is the minimal number density of reactive species at which the formation of lethal lesions is still possible. Processes that accomplish the transport depend on the LET and the shock wave-induced collective flow
plays an important role by saving the reactive species from annihilation.
Function $\sigma(S_e)$ is the cross section of production of a lethal damage in the nucleus.
It depends on LET and also on environmental conditions of the target (e.g., on the concentration of oxygen).
The dependence of $\sigma$ on $S_e$ comes from the number of reactive species hitting the DNA (which is proportional to LET)
and from the range of their propagation (which is nearly proportional to LET as follows from Eq.~(\ref{eq:SW_cond})).
Therefore, in the first approximation, one may write \cite{Surdutovich_2017_EPJD.71.210}:
\begin{equation}
\sigma(S_e) = \xi \, S_e^2 \ ,
\label{eq_111}
\end{equation}
where $\xi = 5.8 \times 10^{-6}$~nm$^4$/eV$^2$ is a coefficient.
Further details of calculation of parameters entering Eqs.~(\ref{eq_01b})--(\ref{eq_111})
can be found in Refs. \cite{Surdutovich_2014_EPJD.68.353, Nano-IBCT_book}.

%%%%%%%%%%%%%%%%%%%%%%%%%%%%%%%%%5

The effect of each ion can be treated independently from others as long as the average distance
between the paths is considerably larger than the radii of tracks.
Typical doses used in ion beam therapy are small \cite{Amaldi_2005_RepProgPhys.68.1861} and the
above condition is satisfied \cite{Surdutovich_2014_EPJD.68.353}.
Then, the average number of lethal lesions per ion traversing distance $z$ through a cell nucleus
is given by a product of $\frac{{\rm d}N_l}{{\rm d}x}$ and the average length of traverse of all
ions passing through a cell nucleus at a given dose,
\begin{equation}
Y_{l} = \frac{{\rm d}N_{l}}{{\rm d}x} \, \bar{z} \, N_{\rm ion}(d) \ ,
\label{eq_04}
\end{equation}
The average number of ions traversing the nucleus, $N_{\rm ion}$, depends on dose, LET
and the size of cell nucleus: $N_{\rm ion} = A_{\rm n} \, d / S_{e}$, where
$A_{\rm n}$ is the cross sectional nuclear area.

Combining these expressions, the number of lethal lesions can be written as \cite{Verkhovtsev_2016_SciRep.6.27654}:
\begin{equation}
Y_{l} = \frac{\pi}{16} \, \sigma_l(S_e) \, N_{\rm g} \frac{d}{S_e} \ ,
\label{eq_05}
\end{equation}
where $N_{\rm g}$ is genome size, equal to 3.2~Gbp for human cells \cite{Alberts_MolBiolCell}.
This expression is obtained by averaging nuclear DNA density over the cell cycle duration.
Knowing $N_{\rm g}$ for a cell line of particular origin and accounting for the chromatin dynamics during the cell cycle,
one can evaluate the number density of chromatin $n_{\rm s}$ (see Ref. \cite{Verkhovtsev_2016_SciRep.6.27654} for details).

The probability of cell inactivation is obtained by subtracting the probability of zero lethal lesions
occurrence from unity ($\Pi_{\rm inact} = 1 - e^{-Y_l}$), and the probability of cell survival is given
by unity less that of cell inactivation.
The logarithm of cell survival probability with a minus sign is then simply given by equation (\ref{eq_05}),
\begin{equation}
- \ln \Pi_{\rm surv} = Y_{l} = \frac{\pi}{16} \, \sigma_l(S_e) \, N_{\rm g} \frac{d}{S_e} \ .
\label{eq_06}
\end{equation}
This expression relates the empirical parameter $\alpha$ of the LQ model to the physical parameters
of the ion projectiles and biological parameters of the target,
\begin{equation}
\alpha = \frac{\pi}{16} \, \frac{\sigma_l(S_e)}{S_e} \, N_{\rm g} \ .
\label{eq:alpha}
\end{equation}

%%%%%%%%%%%%%%%%%%%%%%%%%%%%%%%%%%%%%%%%%%%%%%%%%%%%%%%%%%%%%%%%%%%%%
%%%%%%%%%%%%%%%%%%%%%%%%%%%%%%%%%%%%%%%%%%%%%%%%%%%%%%%%%%%%%%%%%%%%%
\section{Results and discussion}

\subsection{Analysis of cell survival curves}

\begin{figure*}[htb!]
\centering
\includegraphics[width=0.98\textwidth,clip]{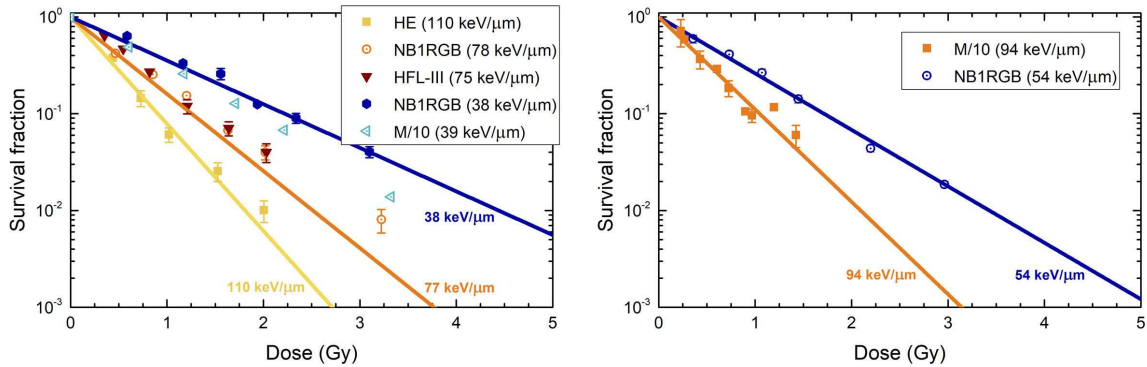}
\caption{Survival curves for several normal tissue human cell lines irradiated with monoenergetic carbon ions:
human embryonic (HE) fibroblast-like cells, skin fibroblast NB1RGB,
normal embryonic lung fibroblast HFL-III, and
M/10 cell line derived from the human mammary epithelial H184B cells.
The survival probabilities calculated at the indicated values of LET are shown with lines.
Experimental data for HE \cite{Suzuki_1996_AdvSpaceRes.18.127},
NB1RGB \cite{Suzuki_2000_IJROBP.48.241, Tsuruoka_2005_RadiatRes.163.494},
HFL-III \cite{Suzuki_2000_IJROBP.48.241} and M/10 \cite{Belli_2008_JRadiatRes.49.597} cells
are shown by symbols.
}
\label{figure_fibroblasts}
\end{figure*}

Figure~\ref{figure_fibroblasts} shows the survival curves for several human normal cell lines
irradiated with monoenergetic carbon ions.
The survival curves calculated using Eq.~(\ref{eq_06}) are shown with lines.
Symbols denote experimental data~\cite{Suzuki_1996_AdvSpaceRes.18.127, Suzuki_2000_IJROBP.48.241,
Tsuruoka_2005_RadiatRes.163.494, Belli_2008_JRadiatRes.49.597} on clonogenic survival of
human embryonic (HE) fibroblast-like cells,
skin fibroblasts NB1RGB, normal embryonic lung fibroblasts HFL-III,
as well as M/10 cells derived from human mammary epithelial cell line H184B.
The figure illustrates the capability of the MSA to reproduce the main trend in the
cellular response to ion-beam irradiation at different values of LET.
These results support the conclusions made in our earlier work \cite{Verkhovtsev_2016_SciRep.6.27654}
where survival of a number of human and rodent cell lines irradiated with protons,
$\alpha$-particles and carbon ions was analyzed.
The results shown in Figure~\ref{figure_fibroblasts} together with our earlier results \cite{Verkhovtsev_2016_SciRep.6.27654}
cover a significant part of radiobiological experiments that have been summarized in the widely known
PIDE database \cite{Friedrich_2013_PIDE}.
On this basis, we confirm further the applicability of the MSA for the description of macroscopic
radiobiological effects
of ion-beam irradiation through understanding of the nanoscale mechanisms of ion-induced biodamage.

In this study we focus on the radiobiological response of normal human cell lines.
It is assumed that the variation of radiosensitivity between the studied cells
(i.e., the variation of survival curves) is rather small and can be neglected
in the first approximation.
The validity of this assumption is justified below.
The variability in radiosensitivity/radioresistance of normal cells of the same origin is much
smaller than that of different tumor cell lines \cite{Suzuki_2000_IJROBP.48.241}.
This can be attributed to more frequent mutations in cancerous cells resulting in
inactivation of specific repair proteins or underexpression of repair enzymes
\cite{Chae_2016_Oncotarget.7.23312}.
A molecular-level understanding of the mechanisms of DNA damage response to ion irradiation
is a complex problem, and we hope that it could be tackled by means of the MSA methodology in our future studies.

Here we consider a number of normal tissue human cell lines and assume that the density of
chromatin and hence the number of complex damage sites do not vary between the different cells.
This allows for the validation of other parameters entering the analytical recipe for the assessment
of ion-induced biodamage presented in Section~\ref{Methods}.
The number density of complex damage sites in the cells is then calculated implying that
the nucleus of a typical human normal cell contains $6.4 \times 10^9$ base pairs \cite{Alberts_MolBiolCell},
and this number is kept constant in the calculations.
However, it is expected that the value of $n_{\rm s}$ may vary significantly between different
tumor cell lines of the same origin.
This may happen because of an abnormal number of chromosomes (so-called aneuploidy)
that is a prominent feature of cancer cells \cite{Thompson_2011_ChromosomeRes.19.433}.
%Aneuploidy is caused by the so-called chromosomal instability which is defined as
%a high rate of loss and gain of whole chromosomes \cite{Lengauer_1997_Nature.386.623}.
As a result, chromosomal numbers in tumor cell lines may vary significantly and differ
from healthy tissue cells where the overall karyotype of the cell population remains diploid
\cite{Suzuki_2000_IJROBP.48.241}.

Figure~\ref{figure_fibroblasts} demonstrates that the assumption made works rather well for the four cell
lines studied.
It should be stressed here that the MSA-based survival curves shown in Figure~\ref{figure_fibroblasts}
were calculated with the same set of parameters described in Section~\ref{Methods}, i.e. without fitting
them for each particular experimental data set as it is commonly done in other radiobiological models
employing the LQ model.
This gives us confidence about the robustness of the MSA-based methodology for predicting cell survival.

\subsection{Dependence of cell survival-related quantities on LET}

\begin{figure}[htb!]
\centering
\includegraphics[width=0.6\textwidth,clip]{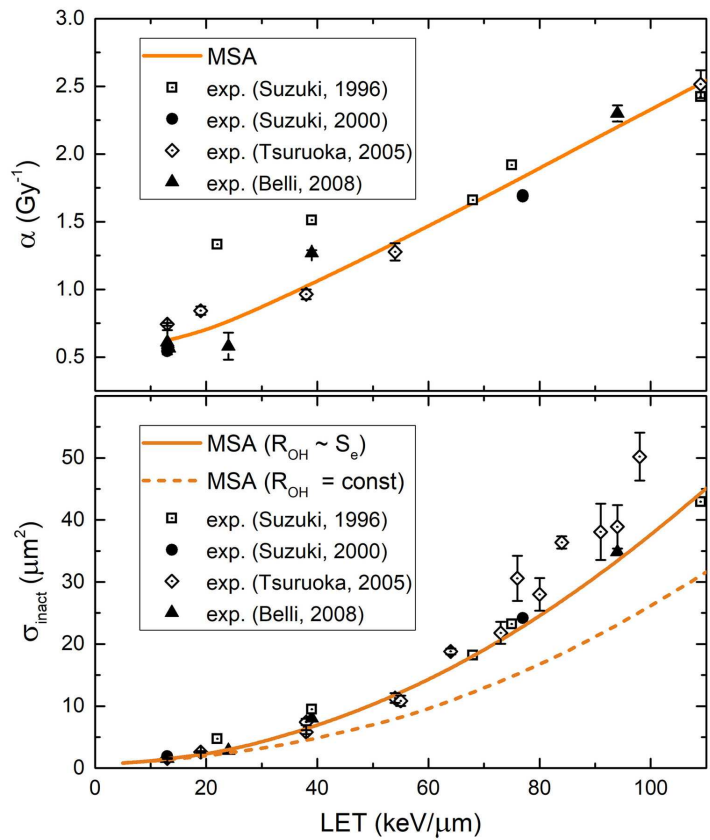}
\caption{Slope of the dose-dependent cell survival curve (i.e., the coefficient $\alpha$ in the LQ model) (upper panel)
and inactivation cross section $\sigma_{\rm inact}$ (lower panel) as functions of LET.
The MSA outcomes (solid lines) are compared with experimental data from
\cite{Suzuki_2000_IJROBP.48.241, Belli_2008_JRadiatRes.49.597, Suzuki_1996_AdvSpaceRes.18.127,
Tsuruoka_2005_RadiatRes.163.494} (symbols).
The dashed line in the lower panel shows the MSA results with a fixed range of reactive species propagation, independent of LET.
}
\label{figure_alpha}
\end{figure}

Apart from comparison with experimental data on clonogenic cell survival, the MSA is utilized
also for the analysis of other radiobiological endpoints and related quantities.
The upper panel of Figure~\ref{figure_alpha} shows the dependence of the slope of survival
curves, $\alpha$, on LET.
The results of MSA-based calculations employing Eq.~(\ref{eq:alpha}) (solid line) are compared with experimental data \cite{Suzuki_2000_IJROBP.48.241, Belli_2008_JRadiatRes.49.597, Suzuki_1996_AdvSpaceRes.18.127,
Tsuruoka_2005_RadiatRes.163.494} (symbols).
Note that the values of $\alpha$ were explicitly given in \cite{Suzuki_2000_IJROBP.48.241, Belli_2008_JRadiatRes.49.597}
only for a few values of LET.
Other experimental figures were obtained from the PIDE database \cite{Friedrich_2013_PIDE}.
In the lowest-LET limit considered in this work (13~keV/$\mu$m) the calculated value $\alpha \approx 0.6$
agrees with the experimental results~\cite{Suzuki_2000_IJROBP.48.241, Belli_2008_JRadiatRes.49.597}.
In the LET range up to about 100 keV/$\mu$m, which is of interest for clinical applications of carbon ions,
$\alpha$ gradually increases and reaches 2.5 at $S_e = 110$ keV/$\mu$m.
%\textcolor{red}{
As follows from Eq.~(\ref{eq:alpha}) an increase of $\alpha$ with LET reflects a non-linear
dependence $\sigma(S_e)$ which is attributed to the indirect mechanism of DNA damage due to the shock wave.
%}
Note that the calculated curve agrees well with an extensive data set compiled from the four different experiments.

We have also analyzed the inactivation cross section $\sigma_{\rm inact}$
which is commonly introduced to describe the effects of charged particle irradiation
in terms of ion fluence $F$ instead of ion dose $d$ \cite{Scholz_chapter_2006}.
The inactivation cross section enters an expression for a fluence-wise definition of cell survival probability,
$-\ln \Pi_{\rm surv} = \sigma_{\rm inact} F$.
Then, using the relation between fluence and dose \cite{Alpen_RadiatBiophys}, $d = F \, S_e / \rho$
(where $\rho$ is the mass density of the medium),
one can calculate the inactivation cross section corresponding to a given level of cell survival as a function of LET,
\begin{equation}
\sigma_{\rm inact}  = -\frac{ \ln \Pi_{\rm surv} \, S_e }{d \, \rho}  \ .
\label{eq_sigma_inact}
\end{equation}

In this study we have analyzed $\sigma_{\rm inact}$ at 37\% survival (this corresponds
to an $e$ times decrease of cell survival probability) and compared with the corresponding
experimental data for the same survival level \cite{Tsuruoka_2005_RadiatRes.163.494}.
The survival probability was calculated using Eq.~(\ref{eq_06}) and the dose %$d_{37}$
corresponding to a 37\% survival level was obtained from this dependence.
Then these values were inserted into Eq.~(\ref{eq_sigma_inact}) to calculate $\sigma_{\rm inact}$
as a function of LET.
This dependence is shown in the lower panel of Figure~\ref{figure_alpha}.
Similar to the above-discussed results, the calculated dependence $\sigma_{\rm inact}(S_e)$ shows
good overall agreement with experimental data.
The inactivation cross section depicted by a solid line was calculated assuming that reactive species
(free radicals and solvated electrons generated due to interaction of the projectile ion and secondary
electrons with water molecules of the medium) are effectively spread away from the ion track via the
ion-induced shock wave.
Such a nanoscale shock wave was predicted in \cite{Surdutovich_2010_PRE.82.051915} and thoroughly
studied, both theoretically and numerically,
in \cite{Surdutovich_2013_SciRep.3.1289, Surdutovich_2017_EPJD.71.285, deVera_2018_EPJD_radiochem}.
According to the outcomes of these studies, the characteristic range of reactive species propagation
increases linearly with LET due to an increasing strength of the shock wave \cite{Surdutovich_2017_EPJD.71.285}.
The dashed line in the lower panel of Figure~\ref{figure_alpha} illustrates the cross section
$\sigma_{\rm inact}$ calculated with a fixed range of reactive species propagation, set to 5~nm.
This value corresponds to a typical range of diffusion-driven propagation of radical species
(mainly OH radicals) in a cellular environment
\cite{Stewart_2011_RadiatRes.176.587, Nikjoo_1997_IJRB.71.467}, which varies in different publications
between 4 and 6~nm.
The OH range of 6~nm was obtained in experiments on X-ray induced DNA strand breaks and cell killing
\cite{Roots_1975_RadiatRes.64.306}, where the average lifetime of OH radicals was estimated
on the order of several nanoseconds.
In the LEM IV model an effective range of different radical species is set to a similar
value of 4~nm \cite{Friedrich_2013_PMB.58.6827}.
In this case the calculated inactivation cross section is systematically smaller than the experimental values.
These outcomes further support the idea that the shock waves induced by ions traversing a biological
medium play a significant role in the indirect mechanisms of ion-induced biodamage on the nanoscale.

\subsection{Evaluation of RBE for different endpoints}
\label{RBE_normal}

\begin{figure*}[htb!]
\centering
\includegraphics[width=0.6\textwidth,clip]{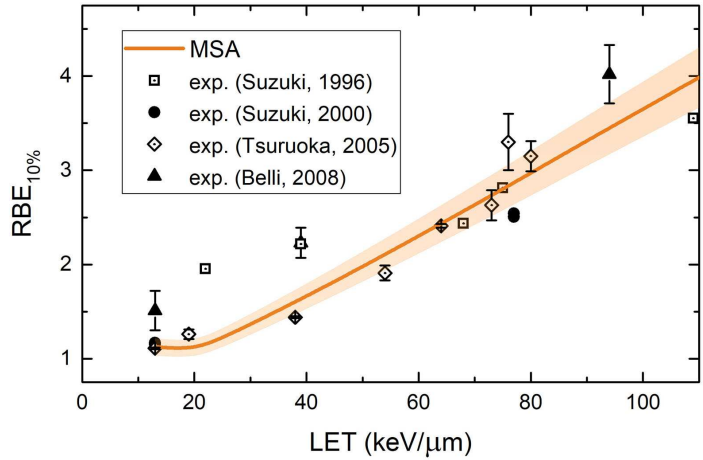}
\caption{RBE at 10\% cell survival for human normal tissue cells irradiated with carbon ions.
Solid line shows the RBE values calculated by means of the MSA.
Shaded area shows a photon dose related uncertainty in RBE due to corresponding
difference in the photon response curves in experiments \cite{Suzuki_1996_AdvSpaceRes.18.127,
Suzuki_2000_IJROBP.48.241, Tsuruoka_2005_RadiatRes.163.494, Belli_2008_JRadiatRes.49.597}.
}
\label{figure_RBE}
\end{figure*}

The practical goal of the phenomenon-based assessment of radiation damage by means of
the MSA is the calculation of RBE.
In the present study, the MSA is applied to evaluate RBE for human normal tissue cell lines
irradiated with carbon ions as an illustrative case study.
Figure~\ref{figure_RBE} shows the dependence of RBE corresponding to a 10\% cell survival on LET.
The RBE$_{10\%}$ is one of the most frequently analyzed endpoints in radiobiological experiments \textit{in vitro}.
As discussed above, we assume that different normal tissue cell lines have similar responses
to ion-beam radiation.
Therefore, for a given value of LET, this response is modeled with a single survival curve.
Despite this simplification, this approach gives reasonable results in agreement with
experimental data as demonstrated in Figures~\ref{figure_fibroblasts} and \ref{figure_alpha}.
To calculate the RBE, the normal tissue cell survival curves obtained by means of the MSA
were normalized to the corresponding photon curves taken from each of the four experiments considered
\cite{Suzuki_1996_AdvSpaceRes.18.127, Suzuki_2000_IJROBP.48.241, Tsuruoka_2005_RadiatRes.163.494, Belli_2008_JRadiatRes.49.597}.
Those reference photon response curves somewhat differ between each other so that the photon dose yielding
a 10\% survival varies between 3.4 and 4.0~Gy.
This variation leads to a dose-related uncertainty in RBE which is illustrated
in Figure~\ref{figure_RBE} by a shaded area.
The solid line shows the RBE$_{10\%}$ values averaged over the four considered experiments.

In the low-LET region ($\sim 13$~keV/$\mu$m) the experimental values of RBE$_{10\%}$ for different normal cells
lie in the range from 1.1 to 1.5, while the RBE$_{10\%}$ increases up to $3.5 - 4.0$ at LET of about $100-110$~keV/$\mu$m.
The calculated values of RBE follow this trend and are in overall good agrement with experimental results.
The calculated RBE matches the experimental figures accounting for the experimental error bars and
the dose-related uncertainty due to averaging over the four different photon curves.

\begin{figure*}[htb!]
\centering
\includegraphics[width=0.98\textwidth,clip]{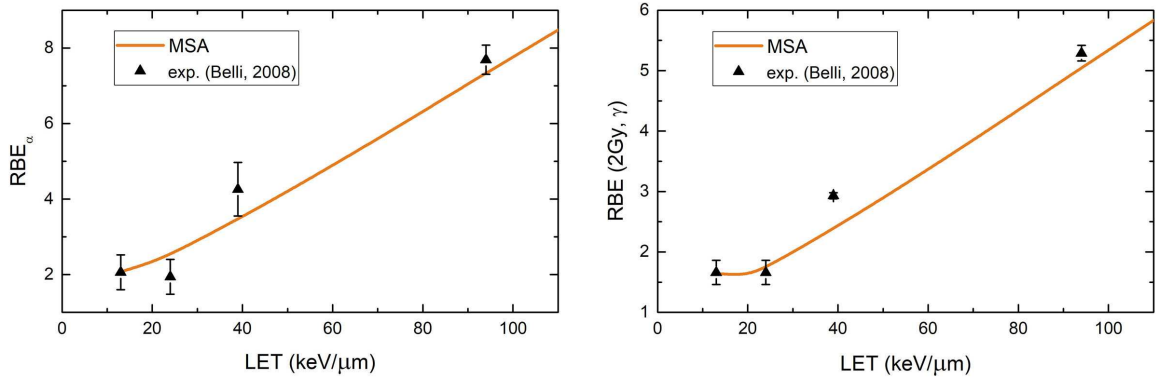}
\caption{RBE for human normal tissue cells irradiated with carbon ions:
RBE$_{\alpha} = \alpha_{\rm ion}/\alpha_{\rm X}$ (left panel) and RBE (2Gy, $\gamma$) (right panel).
Solid lines show the RBE values calculated by means of the MSA.
Symbols denote the experimental data from \cite{Belli_2008_JRadiatRes.49.597}.
}
\label{figure_RBE_other}
\end{figure*}

Besides RBE at 10\% survival, it is rather common to consider other endpoints, such as
RBE at a specific level of cell inactivation (e.g., 50\%, 37\% or 1\%),
RBE$_{\alpha} = \alpha_{\rm ion}/\alpha_{\rm X}$ (which describes the ion biological effectiveness at low doses),
and RBE(2Gy, $\gamma$) that is the RBE at a given ion dose leading to the same inactivation level
as produced by the photon dose of 2~Gy \cite{Kase_2008_PMB.53.37, Belli_2000_IJRB.76.831}.
The latter is regarded as a more relevant endpoint for clinical applications because RBE(2Gy, $\gamma$)
corresponds to the typical dose used in fractionated-dose protocols.
The evaluation of RBE for different endpoints provides a playground to further explore and validate
the predictive power of the MSA.
Figure~\ref{figure_RBE_other} shows the RBE$_{\alpha}$ (left panel) and RBE(2Gy, $\gamma$) (right panel)
for carbon ions as functions of LET.
The calculated curves are compared to the experimental data for M/10 normal cells
from \cite{Belli_2008_JRadiatRes.49.597}, which is the only reference out of the four experiments considered
where the data on RBE for these endpoints has been provided.
For low-LET carbon-ion radiation RBE$_{\alpha} \approx 2$; it then increases
up to about eight in the high-LET region.
%the Bragg peak region.
The RBE(2Gy, $\gamma$) has a similar trend and increases from about 1.6 up to $\sim 5.5 - 6$.
Both specifications of RBE are in overall agreement with the results presented
in \cite{Belli_2008_JRadiatRes.49.597}.

\subsection{Evaluation of RBE at high values of LET}

It is well known from numerous radiobiological experiments with carbon and heavier ions that RBE does not
increase monotonically with LET but has a maximum at $S_e \approx 100-200$~keV/$\mu$m (depending on the ion type)
and gradually decreases at larger $S_e$.
This feature is commonly attributed to the so-called ``overkill'' effect.
The explanation of this effect is that at high LET the energy is deposited into a target cell nucleus by
a small number of ions, and this energy is larger than that needed for cell inactivation.
As a result, such high-LET irradiation produces higher DNA damage than actually required \cite{IBT_book_Linz}.

Different approaches have been adopted in different radiobiological models to account for this effect.
For instance, a ``saturation correction'' due to non-Poisson distribution of lethal lesions in the cell nucleus
was introduced in the LEM and MKM models to describe the radiobiological response for high-LET irradiation
\cite{Hawkins_2003_RadiatRes.160.61, Kase_2008_PMB.53.37}.
Here the analysis presented in the previous section is extended to describe the overkill effect within the MSA framework.

In order to derive the dependence of RBE on LET at large values of stopping power,
let us recall the yield of lethal lesions, $Y_l$, which is defined by Eqs.~(\ref{eq_02})--(\ref{eq_04}).
Combining them, one gets
\begin{equation}
- \ln \Pi_0 = Y_{l} = \xi \, S_e^2 \, \bar{z} \, N_{\rm ion} \ ,
\label{Pi0}
\end{equation}
where $\Pi_0$ is the target cell survival fraction.
Notice, that even though $N_{\rm ion}$ in Eq.~(\ref{Pi0}) is an average number of ions
traversing the cell nucleus, in reality the number of ions is integer.
Therefore, $N_{\rm ion}$ can be redefined as the minimum number of ions required to cause the damage
reflected by the survival fraction of $\Pi_0$.
From Eq.~(\ref{Pi0}) one then derives
\begin{equation}
N_{\rm ion} = \left[ \frac{- \ln \Pi_0}{\xi \, S_e^2 \, \bar{z}} \right] + 1 \ ,
\label{Nion_micro}
\end{equation}
where square brackets denote the integer part of their content.
This expression describes the \textit{sufficient} minimal number of ions required to produce the target biological effect $\Pi_0$.
The dose delivered to the cell nucleus by this number of ions is $d = S_e \, \bar{z} \, N_{\rm ion} / m$,
where $\bar{z}$ is the average length of traverse of an ion through the nucleus and $m$ is the mass of the nucleus.

Let us calculate the RBE for a given biological effect described by a cell survival probability $\Pi_0$.
According to the LQ model, Eq.~(\ref{LQ_model}), the logarithm of a given cell survival probability depends
on the photon dose $d_{\gamma}$ of a reference radiation as
$ - \ln \Pi_0 = \alpha_{\gamma} d_{\gamma} + \beta_{\gamma}d^2_{\gamma}$.
Then the RBE can be calculated by dividing $d_{\gamma}$ by the dose due to ions sufficient to achieve $\Pi_0$.
Using the relations above, this ratio can be written as follows
\begin{equation}
{\rm RBE} = \frac{d_{\gamma}}{ S_e \, \bar{z} \, N_{\rm ion} / m } =
\frac{d_{\gamma}}{ \frac{S_e  \bar{z} }{m} \, \left( \left[ \frac{- \ln \Pi_0}{\xi \, S_e^2 \, \bar{z}} \right] + 1 \right) } \ .
\label{RBE_upd}
\end{equation}
At small values of $S_e$ the integer part is much larger than unity (i.e., the number of ions incident
on a target nucleus $N_{\rm ion} \gg 1$) so that the RBE is a linear function of $S_e$,
independent of $\bar{z}$.
The range of LET where this condition is fulfilled was analyzed above in Section~\ref{RBE_normal}.
At larger values of $S_e$ the integer part gradually approaches zero
so that RBE is asymptotically inversely proportional to $S_e$, i.e. ${\rm RBE} = d_{\gamma} \, m / S_e \, \bar{z}$.

\begin{figure*}[htb!]
\centering
\includegraphics[width=0.6\textwidth,clip]{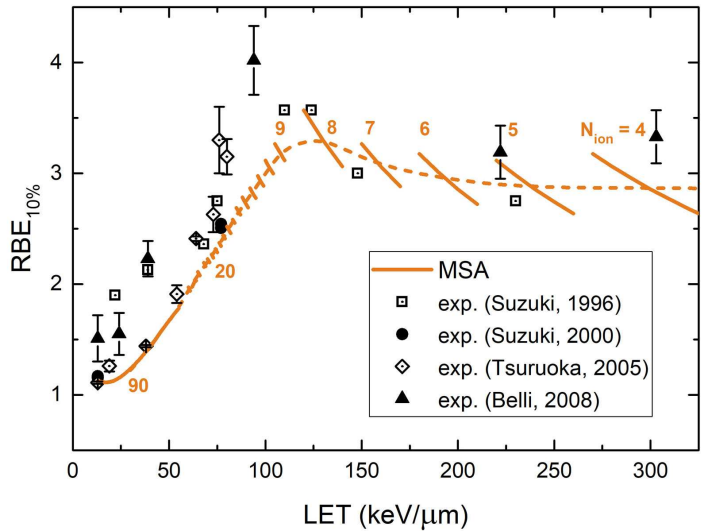}
\caption{RBE at 10\% cell survival for human normal tissue cells irradiated with carbon ions.
The results are obtained by means of Eq.~(\ref{RBE_upd}).
In the high-LET region the RBE becomes inversely proportional to LET, and the absolute values of RBE
depend on the number of ions that traverse a cell nucleus.
The values of $N_{\rm ion}$ corresponding to different segments of the calculated curve
are indicated. The dashed line is a guide to the eye connecting median points of the hyperbolas.
Symbols depict experimental data from \cite{Suzuki_1996_AdvSpaceRes.18.127,
Suzuki_2000_IJROBP.48.241, Tsuruoka_2005_RadiatRes.163.494, Belli_2008_JRadiatRes.49.597}.
}
\label{figure_RBE_overkill}
\end{figure*}

Figure~\ref{figure_RBE_overkill} shows the RBE at 10\% survival in a broad range of LET up to about 300~keV/$\mu$m.
The curves calculated by means of Eq.~(\ref{RBE_upd}) are compared with experimental data described in
Figure~\ref{figure_RBE} and those extended towards the larger values of LET.
As follows from the figure, at small and moderate values of LET (up to about 80~keV/$\mu$m) the dependence
of RBE on LET follows a straight line and corresponds to the results shown in Figure~\ref{figure_RBE}.
Then the RBE starts to deviate from a linear dependence as the number of ions traversing the nucleus becomes
comparable (in the order of magnitude) to 1. With an increase of LET a smaller number of ions is needed to deliver
the dose that would inactivate the cell.
Since the number of ions is an integer quantity,
the RBE($S_e$) dependence transforms into a series of segments of hyperbolas,
where each segment corresponds to a given number $N_{\rm ion}$.
The values of $N_{\rm ion}$ corresponding to different hyperbolas are indicated in the figure.
The dependence depicted shows good overall agreement with experimental data from
\cite{Suzuki_1996_AdvSpaceRes.18.127, Belli_2008_JRadiatRes.49.597} in the range up to $\sim 300$~keV/$\mu$m.
An important observation is that in this regime the dose needed to achieve a given biological effect
is deposited by only a few ions, and the number of ions $N_{\rm ion}$ is a discrete quantity.
The fact that a given number of ions may produce more damage than would be needed for a given biological
effect leads to a significant (up to 20\%) variation in RBE.
To the best of our knowledge, this phenomenon has never been considered in existing radiobiological studies
dealing with the overkill effect.
Therefore, the variation of RBE should be taken into consideration in the analysis of experimental data
on small-dose irradiation with high-LET ions.

\begin{figure*}[htb!]
\centering
\includegraphics[width=0.6\textwidth,clip]{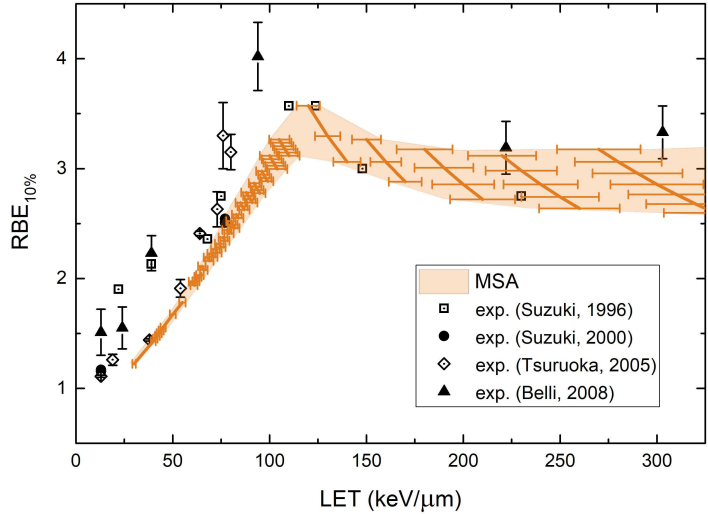}
\caption{RBE at 10\% cell survival for carbon-ion irradiation.
The shaded area shows the variation of RBE with an account for statistical uncertainties in the number of ions
traversing a cell nucleus and the LET.
Symbols depict experimental data from \cite{Suzuki_1996_AdvSpaceRes.18.127,
Suzuki_2000_IJROBP.48.241, Tsuruoka_2005_RadiatRes.163.494, Belli_2008_JRadiatRes.49.597}.
}
\label{figure_RBE_overkill_uncertainty}
\end{figure*}

In \textit{in vitro} experiments with pencil-beam radiation, $N_{\rm ion}$ and $S_e$ are stochastic quantities
that vary randomly within the beam.
To account for this, the above-described analysis can be extended by calculating the variation of RBE (\ref{RBE_upd})
due to statistical uncertainties of $N_{\rm ion}$ and $S_e$.
Typical pencil beams used in proton- or carbon-ion therapy have the lateral size of several millimeters \cite{Philips_Oncology_book}.
Taking as an estimate a typical cell diameter of about $20-50$ $\mu$m, one gets that
$N_{\rm cell} \sim 10^3 - 10^4$ cells will be irradiated by such a beam.
For a large number $N_{\rm cell} \gg 1$ the number of ions hitting a cell can be evaluated as
$N_{\rm ion} = \bar{N}_{\rm ion} \pm \Delta N_{\rm ion}  \approx \bar{N}_{\rm ion} \pm \sqrt{{\bar{N}}_{\rm ion}/N_{\rm cell}}$,
where $\bar{N}_{\rm ion}$ is defined by Eq.~(\ref{Nion_micro}).
For $N_{\rm cell} = 10^3$, the variation $\Delta N_{\rm ion}$ does not exceed 1.5\% which leads to
a minor variation of RBE compared to the data presented in Figure~\ref{figure_RBE_overkill}.
The variation of LET can be extracted from experimental data, e.g.,
the value of $77 \pm 1.8$~keV/$\mu$m ($\Delta S_e = 2.3\%$) was reported in \cite{Suzuki_2000_IJROBP.48.241}.
It is expected that the statistical error for LET will grow with an increase of LET.
However, the publications on irradiation of cells with high-LET carbon ions, which are used for the comparison
in this study, provided only the averaged values of LET and not the uncertainties.
To estimate a magnitude of this variation at high LET, we use the numbers provided in \cite{Dang_2011_EPJD.63.359}
for irradiation of plasmid DNA with carbon ions at the spread-out Bragg peak ($S_e \approx 189$~keV/$\mu$m).
In that paper the experimental uncertainty of LET was about 8\%.
Figure~\ref{figure_RBE_overkill_uncertainty} shows the RBE$_{10\%}$ with the statistical uncertainty in LET
being taken into account.
Based on the available experimental data on $\Delta S_e$, the uncertainty was set to 2.5\% in the low-LET region
(below 100~keV/$\mu$m) and increased gradually up to 8\% at high values of LET.
The uncertainty in LET leads to a broadening of the segments of hyperbolas shown in Figure~\ref{figure_RBE_overkill},
especially in the region of high LET where $\Delta S_e$ is large.
As a result, the RBE as a function of LET %will be not a set of piece-wise functions but rather
transforms into a continuous band which is depicted in Figure~\ref{figure_RBE_overkill_uncertainty} by a shaded area.

\section{Conclusion}

In this study the MultiScale Approach to the physics of radiation damage with ions was applied
to calculate survival probabilities and relative biological effectiveness for human normal cell
lines irradiated with carbon ions at different values of LET.
As a byproduct of this analysis, other radiobiological parameters such as inactivation cross
section were calculated and compared with available experimental data.
Normal cell lines have been chosen as a case study because their proliferation is highly
organized as compared to tumor cells.
This allowed us to test robustness of the MSA-based methodology and validate its key parameters,
e.g., the genome size which remains almost constant in different normal cell lines but may
vary greatly in different tumor cells.
We also successfully tested the hypothesis that the response of different normal cells to ion-beam
irradiation does not vary significantly and thus can be described by a single survival curve for each
value of LET.
Good agreement with a large set of experimental data on clonogenic cell survival, inactivation
cross section and RBE for various endpoints supports our earlier conclusions about the predictive power
of the MSA.
In the current and our earlier analysis of survival probabilities for different cell types irradiated with ions,
we have covered a significant part of radiobiological experiments that are summarized in the Particle
Irradiation Data Ensemble database.
On this basis, we further confirmed the applicability of the MSA for the description
of macroscopic radiobiological effects
of ion-beam irradiation through understanding of the nanoscale mechanisms of ion-induced biodamage.
Finally, the MSA was utilized to describe the ``overkill'' effect which results in a decrease of RBE
at high values of linear energy transfer.
The results obtained are also in good overall agreement with experimental data.
We demonstrated that for a given number of high-LET ions traversing a cell nucleus the RBE becomes inversely
proportional to LET.
The fact that a given number of ions may produce more damage than would be needed for a given biological
effect leads to a significant (up to 20\%) variation in RBE.
We therefore suggest that this effect should be carefully considered in the analysis of experimental data
on small-dose irradiation with high-LET ions as it may lead to re-evaluation of the RBE in the high-LET regime.

%%%%%%%%%%%%%%%%%%%%%%%%%%%%%%%%%%%%%%%%%%%%%%%%%%%%%%%%%%%%%%%%%%%%%

\ack
This work was supported through the DKFZ Postdoctoral Fellowship (granted to AV)
and by the Deutsche Forschungsgemeinschaft.

%%%%%%%%%%%%%%%%%%%%%%%%%%%%%%%%%%%%%%%%%%%%%%%%%%%%%%%%%%%%%%%%%%%%%

\end{document}